\documentclass[aps, prl, twocolumn, groupedaddress, 10pt, showpacs ,superscriptaddress]{revtex4-1}

\usepackage{graphicx}
\usepackage{color}
\usepackage{amsfonts,amsmath,amsthm,amssymb,amscd}
\usepackage{dsfont}
\usepackage{mathrsfs}
\usepackage{mathtools}
\usepackage{yfonts}
\usepackage[dvipsnames]{xcolor}
\usepackage[colorlinks]{hyperref}
\hypersetup{
    citecolor = {blue},
    urlcolor = {teal}
}

\def\ii{{\rm i}}
\newcommand{\dd}{{\rm d}}
\def\bra#1{\mathinner{\langle{#1}|}}
\def\ket#1{\mathinner{|{#1}\rangle}}
\newcommand{\braket}[2]{\langle #1 \vert #2 \rangle}
\newcommand{\RaR}{\mathbb{R}}

\def\ua{\uparrow}
\def\da{\downarrow}

\definecolor{darkred}{rgb}{0.6,0.0,0}

\def\g{{\mu}}
\def\v{{v}}
\def\B{{B}}
\def\lm{{\lambda}}

\begin{document}

\title{From Interacting Particles to Equilibrium Statistical Ensembles}

\author{Enej Ilievski}
\affiliation{Institute for Theoretical Physics Amsterdam and Delta Institute for Theoretical Physics,
University of Amsterdam, Science Park 904, 1098 XH Amsterdam, The Netherlands}

\author{Eoin Quinn}
\affiliation{Institute for Theoretical Physics Amsterdam and Delta Institute for Theoretical Physics,
University of Amsterdam, Science Park 904, 1098 XH Amsterdam, The Netherlands}

\author{Jean-S{\'e}bastien Caux}
\affiliation{Institute for Theoretical Physics Amsterdam and Delta Institute for Theoretical Physics,
University of Amsterdam, Science Park 904, 1098 XH Amsterdam, The Netherlands}

\date{\today}

\begin{abstract}
We argue that a particle language provides a conceptually simple framework for the description of anomalous equilibration
in isolated quantum systems. We address this paradigm in the context of integrable models, which are those with particles that are stable against decay. In particular, we demonstrate that a complete description of equilibrium ensembles for interacting integrable models requires a formulation built from the mode occupation numbers of the underlying particle content, mirroring the case of non-interacting particles. This yields an intuitive physical interpretation of generalized Gibbs ensembles, and reconciles them
with the microcanonical ensemble.
We explain how previous attempts to identify an appropriate ensemble overlooked an essential piece of information, and
provide explicit examples in the context of quantum quenches.
\end{abstract}

\pacs{02.30.Ik,05.70.Ln,75.10.Jm}

\maketitle


The last decade has seen experimental techniques in optical lattices establish themselves as a fruitful playground for testing 
paradigms of quantum statistical physics~\cite{Greiner02,Bloch_review,Polkovnikov_review,LGS16_review}.
A hallmark achievement has been the observation of anomalous equilibration in ultracold bosonic
condensates~\cite{Kinoshita06,Trotzky12,Gring12,Cheneau12,Langen15} where the non-ergodic character of quantum dynamics is attributed 
to the fact that they lie in the vicinity of an integrable 
point~\cite{KLA07,Manmana07,MK08,IC09,CK12,Eisert_review,DeNardis15,Panfil15, 
BD16_review,CC16_review,Cazalilla16_review,DeLuca16_review,EF16_review,IMPZ16_review,VR16_review}.
It became clear that a new framework would be required to properly describe such anomalous relaxation processes.
In this Letter we advocate a shift from conventional interpretations by offering a simple conceptual framework centred around the 
notion of particles. We argue that the consideration of anomalous equilibration in isolated quantum many-body systems naturally leads 
to the study of integrable models since they posses stable particle excitations.
The techniques of integrability in turn allow for a complete description of generalized equilibrium states.

Recent efforts on the subject have predominantly revolved around the notions of pre-thermalization and generalized Gibbs
ensembles (GGE). By invoking locality and entropy extremization, the GGE was introduced 
as the canonical statistical ensembles in which the Hamiltonian is supplemented with an extensive amount of additional conservation
laws stemming from integrability.
Initial studies focused on non-interacting particles, for which the conserved operators forming the GGE are the 
single-particle mode numbers~\cite{RMO06,Rigol07,Cazalilla06,Rigol11,CEF11,CEF12,CEF12II,EEF12,Fagotti13,Sotiriadis14}. Formulating 
the GGE for genuinely interacting models turned out to be more elusive however, due to the less obvious structure of the
local conservation laws~\cite{Fioretto12,Kormos13,Kormos14,FE13,CE13,Bertini14,FCEC14,DeNardis14,WoutersGGE14,PozsgayGGE14,GA14}.

Here we address the problem from a different viewpoint, and exploit the fact that, in contrast to generic (ergodic) dynamical 
systems, integrable systems exhibit stable collective excitations which can be identified as particles. While in non-interacting 
systems particles do not experience mutual collisions, the distinguishing property of interacting integrable models is that particles 
undergo a completely elastic scattering without particle production or decay~\cite{Bethe31,ZZ79}.
In this Letter we convey how anomalous thermalization in integrable systems reflects the existence of particles and use this 
insight to put forward a complete and universal description of local equilibrium states.

It turns out that our perspective demystifies and clarifies the physical picture of the unconventional 
equilibration seen in non-ergodic interacting models, unifying its description with the case of non-interacting theories.
Another appealing aspect of adopting particle interpretation is that it eliminates the need of 
rather technical and subtle concepts introduced in the previous literature, e.g. recently advocated `weaker forms' of locality such as
quasi-locality~\cite{IlievskiGGE15,IMPZ16_review} and semi-locality~\cite{Panfil15}, and makes the logic of
the `truncated GGE'~\cite{Fagotti13} obsolete. These auxiliary concepts, we argue, only undesirably obfuscate
a clean physical picture. The very existence of an extensive hierarchy of higher (local and non-local) conservation laws is in fact 
a direct manifestation of the elastic (completely factorizable) scattering of particles.
It is the particles that are the inherently local objects of integrable theories~\cite{Bethe31}!

In a closed quantum system, relaxation towards local equilibrium is typically formulated as an initial value
problem -- the so-called quantum quench -- where a question of main interest is to characterize
local equilibria which emerge as quasi-stationary states in the unitary relaxation process.
This task can be naturally explained in the particle language, owing to the fact that integrability ensures that particles have 
infinite lifetimes. This implies that the information about the particle content at the initial time gets preserved for arbitrary long 
times. As we shall shortly discuss, this not only suffices to fully characterize the content of the stationary state, but it
also proves necessary.

Prescribing an initial state can be viewed as exciting a macroscopic number of particles which 
subsequently participate in a scattering process. A local equilibrium state emerges dynamically after long times.
These are only meaningfully defined for a thermodynamically large system, which allows for a simplified description due
to two mechanisms:
\emph{dephasing} and \emph{Eigenstate Thermalization Hypothesis (ETH)}.
Dephasing amounts to discarding the information contained in the dynamical phases of eigenstates differing by 
$\mathcal{O}(1/L)$ in energy~\cite{CE13,Caux16_review}, which averages out in the course of the relaxation process. ETH states that
individual microstates which are related by non-extensive modifications of quantum numbers are locally indistinguishable and give the 
same (generalized) free energy and expectation values of local observables~\cite{YY69}.
No distinction between microcanonical and canonical ensembles is necessary as
the former are understood as an unbiased collection of eigenstates which share the same set of 
distributions of quantum numbers, which for thermodynamically large systems exactly matches the canonical description,
with the density operators of the particles providing a labelling of the microcanonical shells~\cite{YY69}.

The task of finding and parametrizing the stationary state after a quench therefore amounts to detect the particle distributions 
from the initial state. There exist various strategies for performing this technical step. For instance, the one pursued in the 
seminal works on interacting quenches was to employ the Quench Action method~\cite{CE13} and compute the leading thermodynamic 
contribution of the overlap coefficients between the initial state and Bethe eigenstates~\cite{DeNardis14,WoutersGGE14,PozsgayGGE14}.
While this has been successfully accomplished for certain special states~\cite{Pozsgay14}, it seems to pose a formidable task for 
generic states. This can be overcome with the approach of Ref.~\cite{StringCharge} which derives exact operator 
representations for $\hat{\rho}_j$, allowing for efficient extraction of the particle densities from initial 
states which admit a (exact or approximate) representation in terms of matrix-product states~\cite{FCEC14,IlievskiGGE15}.

The remaining discussion is devoted to rigorous justification of our claims.
By carefully revisiting the previous formulations we shall demonstrate that our proposal not only unveils the
true character of local conserved quantities, but it also captures a part of the manifold of equilibrium ensembles missed by
previous proposals.
\paragraph{Generalized Gibbs ensembles.--}
The concept of a generalized Gibbs ensemble (GGE) has been proposed, conventionally given in the 
form~\cite{Rigol07,RMO06,Cazalilla06,Rigol11},
\begin{equation}
\hat{\varrho} = \mathcal{Z}^{-1}\exp{\Big[-\sum_{i}\lm_{i}\hat{Q}_{i}\Big]},
\label{eqn:GGE_abstract}
\end{equation}
where $\hat{Q}_{i}$ is some appropriately chosen set of conserved operators, $\lm_{i}$ the associated Lagrange multipliers,
and $\mathcal{Z}={\rm Tr}\,\hat{\varrho}$.
The GGE rests on the `principle of locality', which states that the properties of local observables are completely 
characterized by including all local conserved charges $\hat{Q}_{i}$. This provides a greatly simplified description compared to the 
diagonal ensemble, which retains the entire information about the initial condition.

All proposals to date have attempted to define the GGE in terms of the Hamiltonian and other local charges
obtained from the traditional algebraic Bethe Ansatz procedure~\cite{Faddeev_arxiv,KorepinBook}.
These suffer from two major drawbacks however: (i) they shed no light on the physical interpretation of the charges $\hat{Q}_{i}$,
and (ii) the question whether a given trial set of charges provides a complete and non-redundant characterization of local 
equilibria for a given model remained obscure. In fact we shall show with explicit examples that the set of local charges from the previous 
proposals omits a crucial part of the local information necessary to characterize general equilibrium states.
The shift of perspective which we advocate in this Letter overcomes these difficulties.

As the entire spectrum of an integrable model is characterized in terms of stable particles,
the number operators of these particles provide the most natural complete set for the local conserved
$\hat{Q}_{i}$ in Eq.~\eqref{eqn:GGE_abstract}. Specifically, expressing the ensemble in terms of particle density operators $\hat{\rho}_{j}(v)$,
it takes the form
\begin{equation}
\hat{\varrho} = \mathcal{Z}^{-1}\exp{\Big[- \sum_{j}\int_{\mathbb{R}}\dd \v\;\g_{j}(\v)  \hat{\rho}_{j}(\v)\Big]},
\label{eqn:complete_GGE}
\end{equation}
where the index $j$ runs over distinct particle types~\footnote{A technical note: for models whose Bethe ansatz equations take a 
nested form, only particles associated with primary roots enter the ensemble, as these in turn determine the auxiliary roots.}, $\v$ 
is a rapidity variable which parametrizes their momenta $p_{j}(\v)$, and $\g_{j}(\v)$ are rapidity dependent chemical potentials.
The eigenvalue densities $\rho_{j}$ of the operators $\hat{\rho}_{j}$ completely determine the macrostates
of the system~\cite{YY69}.
Individual microstates drawn from the same macrostate differ only by rearrangements of the occupation of the $L\rho_{j}(\v)\dd \v$ 
quantum numbers in a rapidity window $\dd \v$, and thus have a negligible effect on the local correlations in the
thermodynamic limit. The entropy takes the universal form given explicitly below (cf. Eq.~\eqref{eqn:Yang-Yang_entropy}).
This is understood as the integrable analogue of ETH used to characterize the thermalization of eigenstates in ergodic 
interacting systems~\cite{Deutsch91,Srednicki94,RigolNature,Rigol_review}.

\paragraph{Heisenberg model.--}
We focus on the prototypical integrable model, the Heisenberg spin-$1/2$ chain,
$\hat{H} \simeq -\sum_{i}\vec{S}_{i}\cdot \vec{S}_{i+1}$. The particle content, with respect to a fully polarized state, consists of 
magnons and $j$-particle bound states thereof, with momenta
$p_{j}(\v)=-\ii \log\big((\v+j\tfrac{\ii}{2})/(\v-j\tfrac{\ii}{2})\big)$~\cite{Takahashi71}.

Recently a set of local charges $\hat{X}_{j}(\v)$ for the Heisenberg model has been identified~\cite{IMP15}, which allows 
macrostates to be uniquely determined from the values of charge densities $X_{j}$ via string-charge duality~\cite{StringCharge}.
This takes the form of a discrete wave equation
\begin{equation}
\rho_{j} = \square X_{j},
\label{eqn:wave_equation}
\end{equation}
where $\square$ is a d'Alembertian on $X_{j}(\v)$ defined for a set of functions $f_{j}(\v)$ as
\begin{equation}
\square f_{j} = f^{+}_{j} + f^{-}_{j} - f_{j-1} - f_{j+1},
\label{eqn:box}
\end{equation}
where $f^{\pm}_{j}(\v) = f_{j}(\v \pm \tfrac{\ii}{2} \mp \ii 0^{+})$ and
$0^{+}$ denotes an infinitesimal which acts as a regulator.

Although previous attempts at identifying a complete ensemble were built directly from a discrete basis of charges
obtained from $\hat{X}_{j}$, the strategy they used was to incorporate the knowledge of charge densities to enforce 
constraints on the space of macrostates, thus tacitly bypassing the canonical form of Eq.~\eqref{eqn:GGE_abstract} which is merely a 
formal series.
Since it remains unclear how to unambiguously interpret it, we instead employ a continuous set of local charges $\hat{X}_{j}(\v)$
and cast the most recently proposed GGE \cite{IlievskiGGE15,IMPZ16_review} in the analytic form
\begin{equation}
\hat{\varrho} = \mathcal{Z}^{-1}\exp{\Big[- \sum_{j}\int_{\mathbb{R}}\dd \v\;\lm_{j}(\v) \hat{X}_{j}(\v)-  \B \hat{S}^{z}\Big]},
\label{eqn:GGE_analytic}
\end{equation}
where functions $\lm_{j}(\v)$ are Lagrange multipliers and $\B$ couples to the magnetization $\hat{S}^{z}$.
While this appears reasonable, due to the correspondence between particle's distributions $\rho_{j}$ and charge densities $X_{j}$ in 
Eq.~\eqref{eqn:wave_equation}, it however fails to capture generic equilibrium states.
In the following we carefully examine the structure of equilibrium states, 
highlight the origin of this incompleteness, and provide explicit examples in the context of a quantum quench scenario.
This leads us to the conclusion that the complete description of thermodynamic ensembles necessitates the particle-based formulation of Eq.~\eqref{eqn:complete_GGE}.

\paragraph{Technical background.--}
Integrability of the model provides a commuting family of operators $\hat{T}_{j}$, known as transfer matrices 
(defined in Supplementary Material (SM)~\cite{SM}).
Following refs. \cite{IMP15,IlievskiGGE15,StringCharge}, a set of conserved charges is naturally identified,
\begin{equation}
\hat{X}_{j}(\v) = \frac{1}{2\pi \ii} \partial_{\v}\log \hat{T}_{j}(\v+\tfrac{\ii}{2}).
\end{equation}
These are Hermitian for $\v \in \mathbb{R}$, and local for
$\v \in \mathcal{P}\equiv \{\v\in \mathbb{C}:|{\rm Im}(\v)|<\tfrac{1}{2}\}$, a domain in the complex plane known as the `physical strip'.

The eigenvalue densities $X_{j}$ of the charges  $\hat{X}_{j}$ are expressed in terms of the particle densities compactly as
\begin{equation}
X_{j} = G_{j,k}\star \rho_{k},\quad G_{j,k}=\sum_{m=1}^{j}a_{|j-k|-1+2m},
\label{eqn:string-charge}
\end{equation}
where the kernel  $a_{j}(\v)=-\tfrac{1}{2\pi}\partial_{\v}p_{j}(\v)$, $(f\star g)(\v) \equiv \int_{\RaR}\dd t \,f(\v-t)g(t)$ denotes 
convolution, and repeated indices are summed over. The matrix kernel $G$ is the Green's function of the discrete d'Alembertian
$\square$, $(\square G_{j,k})(\v)=\delta_{j,k}\delta(\v)$. This enables to
invert the relationship between $X_{j}$ and $\rho_{j}$, which yields Eq.~\eqref{eqn:wave_equation}.
The thermodynamics of integrable models is conveniently treated in the language of Thermodynamic Bethe Ansatz
(TBA)~\cite{YY69,Takahashi71,Gaudin71,TS72,HubbardBook}.
We shall proceed with minimal technicality, and refer the reader to the SM and references therein~\cite{SM}.
The partition function is cast as a functional integration over the densities $\rho_{j}$,
\begin{equation}
\mathcal{Z} = \int \mathcal{D}[\rho_{j}]
\exp{\Big[-L\sum_{j}\int_{\mathbb{R}}\dd \v\,\Big(\g_{j}(v) \rho_{j}(v) +\mathfrak{s}_{j}(v)\Big)\Big]},
\nonumber
\end{equation}
where the  combinatorial weight is given by the Yang--Yang entropy density per mode~\cite{YY69},
\begin{equation}
\mathfrak{s}_{j}(\v) = \rho_{j}\log \Big(1+\frac{\bar{\rho}_{j}}{\rho_{j}}\Big) +
\bar{\rho}_{j}\log \Big(1+\frac{\rho_{j}}{\bar{\rho}_{j}}\Big).
\label{eqn:Yang-Yang_entropy}
\end{equation}
The functions $\bar{\rho}_{j}$ represent the densities of unoccupied modes, which obey the Bethe--Yang equations~\cite{YY69,Takahashi71}
\begin{equation}
\rho_{j}+\bar{\rho}_{j}=a_{j}-a_{j,k}\star \rho_{k},
\label{eqn:Bethe-Yang}
\end{equation}
where the integral kernels $a_{j,k}$ are given by  $a_{j,k}=G_{j,k-1}+G_{j,k+1}$.
The dominant contribution to $\mathcal{Z}$ in the thermodynamic limit is given by the saddle-point,
yielding a set of coupled non-linear integral equations known as the  TBA equations~\cite{Zamolodchikov90,Zamolodchikov91},
\begin{equation}
\log Y_{j} =\g_{j} + a_{j,k}\star \log (1+Y^{-1}_{k}),
\label{eqn:canonical_TBA}
\end{equation}
written in terms of $Y$-functions $Y_{j}=\bar{\rho}_{j}/\rho_{j}$. 
Together equations (\ref{eqn:Bethe-Yang}) and (\ref{eqn:canonical_TBA}) completely determine the equilibrium state. They provide the direct relationship between the set of chemical potentials $\g_j$ and the set of macrostates $\rho_j$, demonstrating the completeness of the ensemble given in Eq.~\eqref{eqn:complete_GGE}.

For practical purposes, it is useful to switch to an alternative formulation of the TBA equations. The infinite sum in
Eq.~\eqref{eqn:canonical_TBA} can be eliminated with the use of  kernel identities, yielding a local form
\begin{equation}
\log Y_{j} = d_{j} + s\star \log\big[(1+Y_{j-1})(1+Y_{j+1})\big],
\label{eqn:local_TBA}
\end{equation}
where the source terms are  $d_{j}= \g_{j} - s\star(\g_{j-1}+\g_{j+1})$,  and the convolution kernel  $s(\v)=(2\cosh{(\pi \v)})^{-1}$. 
These are equivalent to Eq.~\eqref{eqn:canonical_TBA} when supplemented with the large-$\v$ asymptotics of the $Y$-functions.
To proceed, care must be taken when inverting the convolution with  $s$, as the \emph{pseudo-inverse} defined through
\begin{equation}
s^{-1}\star f=f^{+} + f^{-},
\end{equation}
has a non-trivial null space, see e.g. \cite{AF09,Tongeren14}.
In particular, while $s^{-1}\star(s\star f)=f$ for any function $f$, in general $s\star(s^{-1}\star f)\neq f$.
We therefore decompose the  source terms as
\begin{equation}
d_{j}= s\star{\lm}_{j}+d^{\emptyset}_{j},
\label{eqn:splitting}
\end{equation}
where ${\lm}_{j}=s^{-1}\star d_j$, and the  `singular components' $d^{\emptyset}_{j}$ are annihilated by $s^{-1}$.
Applying $s^{-1}$ to Eq.~\eqref{eqn:local_TBA} and exponentiating, yields the modified Y-system relations
\begin{equation}
Y^{+}_{j}Y^{-}_{j} = e^{\lm_{j}}(1+Y_{j-1})(1+Y_{j+1}).
\label{eqn:Y}
\end{equation}
In the process the singular components are not lost, but are instead
encoded in the analytic data of the $Y$-functions: their poles and zeros in the physical strip $\mathcal{P}$.
Equation~\eqref{eqn:local_TBA} is reobtained by convolving the logarithm of Eq.~\eqref{eqn:Y} with $s$, after multiplying out the 
singularities $\xi$ with the functions~\cite{KP92,BH11,Tongeren14}
\begin{equation}
t(\v;\xi) =  \tanh{(\tfrac{\pi}{2}(\v-\xi))},\quad |{\rm Im}(\xi)|<\tfrac{1}{2},
\end{equation}
which satisfy $t^{+}t^{-}=1$. As a result the singular components are of the form
\begin{equation}
d^{\emptyset}_{j}(\v) = \sum_{a} \log t(\v;\xi^{j,\rm z}_a)-\sum_{b}\log t(\v;\xi^{j,\rm p}_b),
\label{eqn:singular_components}
\end{equation}
where $\{\xi^{j,\rm z}_a\}$ and $\{\xi^{j,\rm p}_b\}$ are respectively the sets of zeros and poles of
the $Y$-function $Y_{j}$ in $\mathcal P$.

We now address the question of the generality of the charge-based GGE of Eq.~\eqref{eqn:GGE_analytic}.
Using Eq.~\eqref{eqn:string-charge}, this ensemble takes the form of Eq.~\eqref{eqn:complete_GGE} with
$\g_{j} = G_{j,k}\star \lm_{k} +\B j$. Na\"{i}vely the two ensembles could then be assumed equivalent. The crucial point however is 
that the Lagrange multipliers $\lm_j$ are blind to the singular components $d_j^{\emptyset}$. In fact, as $\square \g_{j}=\lm_j$,  the 
Lagrange multipliers precisely match the functions $\lm_j$ introduced in the decomposition of Eq.~\eqref{eqn:splitting}.

Put another way, the d'Alembertian $\square$ inherits a null space from $s^{-1}$, and $\square \g_{j}=\lm_{j}$  demonstrates that the
Lagrange multipliers $\lm_{j}$ of Eq.~\eqref{eqn:GGE_analytic} cannot encode the components of $\g_{j}$ in this null space.
In the following section we provide evidence that singular component contributions  are a generic feature of equilibrium states.
To conclude, it is worthwhile to stress that the null space of $\square$  does not cause a problem for the string-charge duality 
relations, Eqs.~\eqref{eqn:wave_equation} and \eqref{eqn:string-charge}, as the densities $\rho_{j}$ on which $\square$ acts already 
provide the full description of a macrostate.

\paragraph{Examples.--}
We now explicitly demonstrate this limitation of the charge-based GGE of 
Eq.~\eqref{eqn:GGE_analytic} with some examples in the context of a quantum quench. Starting from an initial state $\ket{\Psi}$, the 
equilibrium macrostate $\rho_{j}$ reached in the long-time limit is determined from the expectation values of the particle density operators
\begin{equation}
\rho_{j}(\v) = \lim_{L\to \infty}L^{-1}\bra{\Psi}\hat{\rho}_{j}(\v)\ket{\Psi}.
\label{eqn:exp_rho}
\end{equation}
As described above, it is convenient to encode a macrostate in the $Y$-functions, $Y_{j}=\bar{\rho}_{j}/\rho_{j}$. The condition that the charge-based GGE is a valid ensemble is the absence of poles 
and zeros of these functions in the physical strip $\mathcal P$.

The best studied quenches are those from  (i) the `dimer' state
$\ket{\rm D} = \frac{1}{2^{L/2}}(\ket{\ua \da}-\ket{\da \ua})^{\otimes L/2}$, and (ii) the N\'eel state
$\ket{\rm N} = \ket{\ua \da}^{\otimes L/2}$,  for which the $Y$-functions are explicitly known~\cite{Brockmann14,Mestyan15}.
For the dimer quench these are~\cite{StringCharge}
\begin{equation}
Y_{j}(\v) = \frac{j(j+2)\v^{2}}{(\v+(j+1)\tfrac{\ii}{2})(\v-(j+1)\tfrac{\ii}{2})}.
\label{eqn:Y_dimer}
\end{equation}
The  double zero at the origin immediately indicates that the Lagrange multipliers $\lm_j$ in the charge-based GGE are not sufficient 
to capture it. Indeed, the source terms $d_{j}(\v) = \log \tanh(\tfrac{\pi}{2}\v)^{2}$, determined through Eq.~\eqref{eqn:local_TBA}, 
are in the null space of $s^{-1}$, from which it follows that $\lm_j$=0! 
We stress that the state is  by no means close to the infinite-temperature 
Gibbs state in terms of local correlation functions. Previous works mistakenly assigned non-trivial $\lm_j$ for this
state~\cite{Brockmann14,Mestyan15,IlievskiGGE15}, resulting in the incorrect interpretation of the GGE. The important point is that 
due to the non-trivial null space of $s^{-1}$ it is \emph{not} permissible to define $\lm_{j}$  via Fourier transform, 
namely $\mathcal{F}[\lm_{j}] \neq \mathcal{F}[d_{j}]/\mathcal{F}[s]$.

The N\'eel state is similar. Again the source terms $d_{j}(\v)=(-1)^{j+1}\log \tanh(\tfrac{\pi}{2}\v)^{2}$  consist solely of singular 
components, indicating $\lm_j=0$ for this state also. Indeed, these two examples are members of a particular class of initial states, 
whose equilibrium states can be cast as the partition function of a vertex model for which the initial state provides an 
\emph{integrable} boundary~\cite{Pozsgay13}. Generic initial states do not admit such a description.

We thus also consider more general product states. A complete analytic treatment is now out of scope, as the complexity of
$\rho_{j}$, computed via Eq.~\eqref{eqn:exp_rho}, grow quickly with $j$.
Nevertheless, the lowest $Y$-functions are straightforwardly determined.
A particular example is the state composed of alternating $2$-site domain walls,
$\ket{\uparrow \uparrow \downarrow \downarrow}^{\otimes L/4}$. Explicit expressions are unwieldy, see SM~\cite{SM}.
They indicate non-trivial $\lm_j$, and that $Y_1$ has four zeros at 
$\pm(0.382\pm0.234\, \ii)$ and two poles at $\pm0.155\, \ii$. This analytic data indicates that again the resulting equilibrium state 
is not captured by the GGE of Eq.~\eqref{eqn:GGE_analytic}.

\paragraph{Conclusion.--}
We have shown that a complete description of equilibrium states in interacting integrable models
requires the ensemble to be constructed from the number operators of the model's particle content. 
Any attempt to reduce the description inevitably  leads to a  loss of information, rendering a part of the equilibrium manifold 
inaccessible. The proposal naturally extends the established framework for non-interacting particles to the interacting integrable 
regime. The effect of non-trivial scattering in the interacting case is that  excitations about an equilibrium state get non-trivially 
dressed by the correlations of the state, see e.g. \cite{YY69,QuinnFrolov13}.


Our hope is that the particle-based perspective which we advocate here creates a platform for the
study of equilibration as one leaves the integrable points. The resulting loss of factorized scattering induces
timescales up to which the particle-based description of ensembles may be expected to accurately capture statistical properties of 
quasi-stationary states, the so-called pre-thermalized regime~\cite{KWE11,SilvaPRL13,BertiniPRL15,BertiniJSTAT15}.

\paragraph*{Acknowledgements.}
We thank M. Brockmann, J. De Nardis for valuable comments on the earlier version of this manuscript,
and L. Vidmar for sharing some insightful remarks. E.~I. acknowledges support by VENI grant number 680-47-454 by the Netherlands 
Organisation for Scientific Research (NWO). The authors acknowledge support from the Foundation for Fundamental Research on Matter 
(FOM) and from the Netherlands Organization for Scientific Research (NWO). The work of  J.-S. C. was supported by NWO VICI grant 
number 680-47-605.

\bibliography{Equilibrium_nice}

\setcounter{equation}{0}
\makeatletter
\renewcommand{\theequation}{S\arabic{equation}}

\newpage 

\begin{widetext}
\section{Supplementary Material:\\
From interacting particles to equilibrium statistical ensembles}

This Supplementary Material provides the essential technical background of the Bethe Ansatz framework.
Special attention is devoted to the derivation of the TBA equations, along with the corresponding functional relations and physical 
interpretation of its analytic input. We provide explicit solutions of two widely studied examples of quantum quenches,
the antiferromagnetic N\'eel state and the `dimer state', and outline how to treat generic states. We conclude by making a connection 
to the Quench Action approach~\cite{CE13,Caux16_review}.

\section{String hypothesis and Bethe--Yang equations}

Any eigenstate of the isotropic Heisenberg spin chain of finite length $L$ is associated a unique set of $M$ rapidities.
These are obtained from solutions to Bethe quantization conditions,
\begin{equation}
e^{\ii p(\v)L}\prod_{k=1}^{M}S_{1,1}(\v-\v_{k}) = -1,\qquad \v = \v_{1},\v_{2},\ldots, \v_{M},
\label{eqn:counting_function}
\end{equation}
where $M/L$ is the magnetization density, $p(\v)=-\ii\log\big((u+\tfrac{\ii}{2})/(u-\tfrac{\ii}{2})\big)$,
and the scattering amplitude $S_{1,1}$ is a member of a complete set of scattering amplitudes
\begin{equation}
S_{j}(\v) = \frac{\v - j\tfrac{\ii}{2}}{\v+j\tfrac{\ii}{2}},\quad
S_{j,k} = S_{|j-k|}S_{j+k}\prod_{m=1}^{{\rm min}(j,k)-1}S^{2}_{|j-k|+2m}.
\label{eqn:scattering_amplitudes}
\end{equation}
In the \emph{thermodynamic limit} ($L\to \infty$, $M\to \infty$, keeping $M/L$ fixed) the solutions (rapidities) 
$\v_{j}$, $j=1,2,\ldots M$ (referred to as the Bethe roots) in the complex plane organize in a special way and permit to partition the 
entire spectrum in terms of particles.
A central feature of generic integrable models is the formation of \emph{bounds states},
representing stable particles which exhibit elastic scattering without particle production.
In terms of solutions $\v_{j}$ ($j=1,\ldots,M$) to Eq.~(\ref{eqn:counting_function}), these comprise of rapidities with
non-zero imaginary parts which share common real parts.
According to the \emph{string hypothesis}~\cite{Takahashi71,Gaudin71,TS72}, the spectra of a thermodynamically large
system can be partitioned in terms of complex-valued solutions, which physically represent multi-magnonic excitations
called the strings. A $k$-string solution ($k\in \mathbb{N}$) centred at $\v^{k}_{\alpha}\in \mathbb{R}$ is parameterized as
\begin{equation}
\left\{\v^{k,i}_{\alpha}\right\} = \left\{\v^{k}_{\alpha}+(k+1-2i)\tfrac{\ii}{2}:i=1,2,\ldots k\right\},
\end{equation}
suppressing exponentially small deviations in system size $L$.

In the $L\to \infty$, the string centers become dense on real axis and condense.
This allows to introduce distributions of Bethe $j$-strings $\rho_{j}$ and their hole counterparts $\bar{\rho}_{j}$
(unoccupied solutions allowed by Bethe quantization condition).
In the thermodynamic limit the quantization condition \eqref{eqn:counting_function} gets replaced by linear integral equations which 
are customary called the Bethe--Yang equations~\cite{YY69,Takahashi71,Gaudin71},
\begin{equation}
\rho_{j}+\bar{\rho}_{j} = a_{j} - a_{j,k}\star \rho_{k},
\label{eqn:Bethe-Yang_equations}
\end{equation}
where here and below we assume summation convention over repeated indices, and
introduce convolution operation, $(f\star g)(\v) = \int_{\RaR}\dd t f(\v-t)g(t)$.
The integral kernels $a_{j}$ and $a_{j,k}$ represent the `kinematic data',
i.e. encode full scattering data among all distinct types of strings.
Specifically, they are the derivatives of scattering phase shifts,
\begin{equation}
a_{j}(\v) = \frac{1}{2\pi \ii}\partial_{\v}\log S_{j}(\v),\quad a_{j,k}(\v) = \frac{1}{2\pi \ii}\partial_{\v}\log S_{j,k}(\v).
\label{eqn:scattering_kernels}
\end{equation}

\section{Local charges and thermodynamic spectra}
In accordance with the standard practice in Yang--Baxter integrable models~\cite{Faddeev_arxiv,KorepinBook},
we introduce an infinite set of quantum transfer operators,
\begin{equation}
\hat{T}_{j}(\v) = {\rm Tr}_{\mathcal{V}_{j}}\hat{L}^{(1)}_{j}(\v)\hat{L}^{(2)}_{j}(\v)\cdots \hat{L}^{(L)}_{j}(\v),
\label{eqn:transfer_operators}
\end{equation}
defined as traces over $L$-fold spatially ordered products of Lax operators $\hat{L}^{(i)}_{j}(u)$ acting in
the tensor product $\mathcal{H}\otimes \mathcal{V}_{j}$, where $\mathcal{H}\cong \mathcal{V}^{\otimes L}_{1}$ is the
Hilbert space of the spin chain. Specifically, Lax operators take the form
\begin{equation}
\hat{L}^{(i)}_{j}(\v) = \mathds{1}^{\otimes (i-1)}_{1}\otimes \hat{L}_{j}(\v)  \otimes \mathds{1}^{\otimes (L-i)}_{1},\qquad
\hat{L}_{j}(\v) = \v\,\mathds{1}_{1}\otimes \mathds{1}_{j} + \ii \sum_{\alpha=\{x,y,z\}}\hat{S}^{\alpha}_{1}\otimes \hat{S}^{\alpha}_{j},
\label{eqn:Lax_operator}
\end{equation}
where $\hat{S}^{\alpha}_{j}$ denotes spin-$j/2$ operators acting in $\mathcal{V}_{j}$ which enclose
the algebraic relations $[\hat{S}^{a}_{j},\hat{S}^{b}_{j}]=\ii \epsilon_{abc}\hat{S}^{c}_{j}$.

Conserved operators $\hat{T}_{j}$ mutually commute, $[\hat{T}_{j}(\v),\hat{T}_{j^{\prime}}(\v^{\prime})]=0$
(for all values of $j,j^{\prime}\in \mathbb{N}$ and $\,\v^{\prime}\in \mathbb{C}$), as implied by the Yang--Baxter relation.
A set of local charges is then defined through their logarithmic derivatives~\cite{IMP15,IlievskiGGE15},
\begin{equation}
\hat{X}_{j}(\v) = \frac{1}{2\pi \ii}\partial_{\v}\log \frac{\hat{T}^{+}_{j}(\v)}{\phi^{[+j]}(\v)},
\label{eqn:local_charges_definition}
\end{equation}
where $\phi=T^{+}_{0}$ and the spectral parameter $\v$ should be restricted to the \emph{physical strip} $\mathcal{P}$, defined
as the strip
\begin{equation}
\mathcal{P}=\{\v \in \mathbb{C}:|{\rm Im}(\v)|<\tfrac{1}{2}\}.
\label{eqn:PS}
\end{equation}
We note that the conserved operators $\hat{X}_{j}(\v)$ are only defined on $\mathcal{P}$ (cf.~\cite{IMPZ16_review}), and
become singular as the boundary is approached.
The Heisenberg spin chain Hamiltonian $H$ is proportional to the charge $\hat{X}_{1}(0)$.
By convention we adopt $\hat{H}\equiv \hat{X}_{1}(0)$.

The action of $\hat{X}_{j}$ on Bethe eigenstates in the $L\to \infty$ limit has been obtained in 
refs.~\cite{IlievskiGGE15,StringCharge}. Expressed in terms of Bethe root densities $\rho_{j}$ they take the compact form
\begin{equation}
X_{j} = G_{j,k}\star \rho_{k},
\label{eqn:charge-string_SM}
\end{equation}
with
\begin{equation}
G_{j,k} = \sum_{m=1}^{{\rm min}(j,k)}a_{|j-k|-1+2m}.
\label{eqn:G_definition}
\end{equation}
Inverting relation \eqref{eqn:charge-string_SM} amounts to find an operator $\square$ such that
$(\square G_{j,k})(\v)=\delta_{j,k}\delta(\v)$. To this end we employ kernel identities~\cite{Takahashi71}
\begin{equation}
a_{j} - I_{j,k}s\star a_{k} = s\;\delta_{j,1},
\label{eqn:fusion_with_s}
\end{equation}
where the convolution kernel reads,
\begin{equation}
s(\v) = \frac{1}{2\cosh{(\pi \v)}},
\label{eqn:s-kernel}
\end{equation}
and the incidence matrix $I_{j,k}\equiv \delta_{j,k-1}+\delta_{j,k+1}$ expresses how distinct types of particles interact
with each other. As Eq.~\eqref{eqn:fusion_with_s}, special attention has to be paid to the pole structure of $a_{1}$.
To properly account for the singularity at the boundary of $\mathcal{P}$ we introduce a pseudo-inverse
\begin{equation}
s^{-1}\star f = f^{+} + f^{-},
\label{eqn:left_inverse}
\end{equation}
where
\begin{equation}
f^{\pm}(\v) \equiv f(\v \pm \tfrac{\ii}{2} \mp \ii 0^{+}),
\label{eqn:shifts}
\end{equation}
for some positive infinitesimal $0^{+}$. It has to be stressed that $s^{-1}$ is only a left inverse of $s$,
$s^{-1}\star (s\star f)=f$, while in general there exist functions $f$ for which $s\star (s^{-1}\star f)\neq f$.
This means that $s^{-1}$ has a non-trivial null space~\cite{AF09,Tongeren14}. With aid of the pseudo-inverse we
define a discrete d'Alembertian $\square$, defined on a set of functions $f_{j}$ as
\begin{equation}
\square f_{j} = s^{-1}\star f_{j} - I_{j,k}f_{k} \equiv f^{+}_{j}+f^{-}_{j}-f_{j-1}-f_{j+1}.
\end{equation}
The action of $\square$ on the scattering kernels gives
\begin{equation}
\square a_{j} = \delta_{j,1}\delta,\qquad \square a_{j,k} = I_{j,k}\delta.
\end{equation}
It is now clear from definition \eqref{eqn:G_definition} that
$G_{j,k}$ represents the Green's function of $\square$,
\begin{equation}
\square G_{j,k} = \delta_{j,k}\delta.
\label{eqn:Green_function}
\end{equation}
Relation \eqref{eqn:charge-string_SM} can now be readily inverted by applying $\square$ on both sides, yielding
\begin{equation}
\rho_{j} = \square X_{j}.
\label{eqn:string-charge_SM}
\end{equation}
The distributions of holes $\bar{\rho}_{j}$ can be obtained from Bethe--Yang equations \eqref{eqn:Bethe-Yang_equations} and read
\begin{equation}
\bar{\rho}_{j} = a_{j} - s^{-1}\star X_{j}.
\end{equation}
Before proceeding we wish emphasize that the identification \eqref{eqn:string-charge_SM} capture two vital ingredients for
a proper formulation of complete set of equilibrium ensembles, that is (i) the importance of the physical strip $\mathcal{P}$
and (ii) the role of the regulator in the definition \eqref{eqn:shifts}.


\section{Canonical and local form of TBA equations}
We now turn our attention to the statistical ensemble
\begin{equation}
\hat{\varrho} = \mathcal{Z}^{-1}\exp\Big[-\sum_{j}\int_{\mathbb{R}}\dd \v\, \mu_{j}(\v)\hat{\rho}_{j}(\v)\Big],
\label{eqn:ensemble}
\end{equation}
where the normalization $\mathcal{Z}={\rm Tr}\,\hat{\rho}$ represent the partition sum.
A standard approach to perform thermodynamic considerations is to work under the string hypothesis and introduce a discrete set
of $Y$-functions,
\begin{equation}
Y_{j} = \frac{\bar{\rho}_{j}}{\rho_{j}},
\end{equation}
given as ratios of hole and particle densities for each string species.
In the thermodynamic limit (by accounting for the Yang--Yang entropy as described in the text)
a functional representation for the partition sum yields an infinite set of coupled non-linear integral equations
\begin{equation}
\log Y_{j} = \g_{j} + a_{j,k}\star \log(1+Y^{-1}_{k}).
\label{eqn:canonical_TBA_equations}
\end{equation}
Introducing the inverse of the matrix kernel $(a+1)$, defined via
\begin{equation}
(a+1)^{-1}_{j,k}\star f_{k}= f_{j} - I_{j,k}s\star f_{k},
\end{equation}
enables to disentangle equations \eqref{eqn:canonical_TBA_equations} into a locally-coupled form
\begin{equation}
\log Y_{j} = d_{j} + I_{j,k}s\star \log(1+Y_{k}),
\label{eqn:local_TBA_equations}
\end{equation}
referred to as the local TBA equations. The source terms $d_{j}$ are related to the chemical potentials
\begin{equation}
d_{j} = \g_{j} - I_{j,k}s\star \g_{k}.
\end{equation}
A subtle point here is that any constant term in $\mu_{j}$ does not enter in $d_{j}$ as $1\star s=\tfrac{1}{2}$.
Equation \eqref{eqn:canonical_TBA_equations} can be reobtained by casting Eq.~\eqref{eqn:local_TBA_equations} as
\begin{equation}
\log \widetilde{Y}_{j} - I_{j,k}s\star \log(\widetilde{Y}_{k}) = I_{j,k}s\star \log(1+Y^{-1}_{k}),
\end{equation}
where $\widetilde{Y}_{j}(\v)=Y_{j}(\v)/Y_{j}(\infty)$, and convolving with kernel $(a+1)$.

\subsection{Modified Y-system}
Equations \eqref{eqn:local_TBA_equations} represent a set of coupled non-linear integral equations defined on the real line.
Equivalently, these can be analytically continued in the complex rapidity plane, resulting in an algebraic form
a set of functional relations
\begin{equation}
Y^{+}_{j}Y^{-}_{j} = e^{\lm_{j}}(1+Y_{j-1})(1+Y_{j+1}),
\label{eqn:modified_Y}
\end{equation}
which we refer to as the \emph{modified} Y-system (we stress that shifts on the left hand side are defined with
the prescription given by Eq.~\eqref{eqn:shifts}).

We now exhibit the equivalence between Eq.~\eqref{eqn:modified_Y}
and TBA equations \eqref{eqn:local_TBA_equations}.
Since $Y$-functions of generic equilibrium states are meromorphic functions inside the physical strip $\mathcal{P}$,
a na\"{i}ve `integration' of the (modified) Y-system Eq.~\eqref{eqn:modified_Y} into TBA equations by virtue of Cauchy 
theorem is not possible. Singular parts of $\log Y_{j}$ can nevertheless be easily remedied.
Let us suppose that $Y$-functions $Y_{j}$ posses a set of zeros and poles in $\mathcal{P}$, located at $\{\xi^{\rm z}_{j,a}\}$ and
$\{\xi^{\rm p}_{j,b}\}$, respectively, with the large-$\v$ asymptotics $Y^{(\infty)}_{j}$. Using functions
\begin{equation}
t(\v;\xi) = \tanh{\left(\tfrac{\pi}{2}(\v-\xi)\right)},\quad {\rm Im}(\xi)<\tfrac{1}{2},
\label{eqn:zero_modes}
\end{equation}
which satisfy the property $t^{+}t^{-}=1$, we introduce a set of renormalized $Y$-functions $\widetilde{Y}_{j}(\v)$,
\begin{equation}
\widetilde{Y}_{j}(\v) = \frac{Y_{j}(\v)}{Y^{(\infty)}_{j}}\frac{\prod_{b}t(\v;\xi^{j,\rm p}_{b})}{\prod_{a}t(\v;\xi^{j,\rm z}_{a})},
\end{equation}
which still satisfy the same same modified Y-system \eqref{eqn:modified_Y} but are now analytic in $\mathcal{P}$ and have
asymptotic behaviour $\lim_{|\v|\to \infty}\widetilde{Y}_{j}(\v)=1$. Now convolving with $s$ after taking the logarithm gives
back the local TBA equations,
\begin{equation}
\log Y_{j} = d^{\emptyset}_{j} + s\star \lm_{j} + I_{j,k}s\star \log(1+Y_{k}),
\label{eqn:TBA_full_form}
\end{equation}
where we introduced the splitting of the source terms $d_{j}=s\star \lm_{j}+d^{\emptyset}_{j}$, with the singular part
\begin{equation}
d^{\emptyset}_{j} = \sum_{a}\log t(\v;\xi^{j,\rm z}_{a}) - \sum_{b}\log t(\v;\xi^{j,\rm p}_{b}).
\label{eqn:singular_part}
\end{equation}
It is worthwhile remarking that the physical solutions correspond to real-valued $Y_{j}(\v)$ on the real axis,
requiring the singularities $\xi$ to always appear in complex-conjugated pairs.

Going in the opposite direction, i.e. transforming Eq.~\eqref{eqn:TBA_full_form} is straightforwardly obtained by applying
$s^{-1}$ to both sides and subsequently taking the exponent. In the process the information from the
singular part $d^{\emptyset}_{j}$ gets transferred into the analytic structure of $Y$-functions, while $\lm_{j}$ enter
the modified Y-system \eqref{eqn:modified_Y} as the non-universal (i.e. node-dependent) part.
Notice that setting $\lm_{j}\equiv 0$ recovers the universal form which is common 
in the literature, governing states whose physical input (TBA source terms) consist solely from the null space components.

\paragraph{Gibbs equilibrium.}
The cleanest examples of a state which fulfils the modified set of functional relations \eqref{eqn:modified_Y} is the canonical
Gibbs equilibrium, corresponding to bare dispersions $\g^{\rm Gibbs}_{j}=\beta\,a_{j}$.
Due to absence of zero modes this is equivalent to have $\lambda^{\rm Gibbs}_{j}=\g^{\rm Gibbs}_{j}$, with
$\lambda_{j}(\v)=\beta\,\delta_{j,1}\delta(\v)$, restoring the well-known
TBA source term $d^{\rm Gibbs}_{j}=s\star \lambda^{\rm Gibbs}_{j}=\beta\,s$.

\section{Exact solutions of quantum quenches}

Below we explain how to detect zero modes given a macrostate $\rho^{\Psi}_{j}$. In the context of a quench protocol $\rho^{\Psi}_{j}$ 
can be understood as equilibrium states of a quench with the initial condition $\ket{\Psi}$.
We first consider two simple product states which have represented toy examples in previous studies of quantum quenches in the 
Heisenberg model~\cite{Pozsgay13,FCEC14,WoutersGGE14,PozsgayGGE14,Brockmann14,Mestyan15,IlievskiGGE15,StringCharge}.

\paragraph{Dimer state.}
We first consider the so-called dimer state
\begin{equation}
\ket{\rm D} = \frac{1}{2^{L/2}}\left(\ket{\uparrow \downarrow}-\ket{\downarrow \uparrow}\right)^{\otimes L/2}.
\label{eqn:dimer_state}
\end{equation}
The associated macrostate $\rho_{j}$ expressed in terms of $Y$-functions reads~\cite{StringCharge}
\begin{equation}
Y_{j}(\v) = \frac{j(j+2)\v^{2}}{(\v + (j+1)\tfrac{\ii}{2})(\v - (j+1)\tfrac{\ii}{2})}.
\label{eqn:dimer_solution}
\end{equation}
These can be quickly verified to obey the standard form of Y-system relations~\cite{Zamolodchikov91,KN92,KNS94}
\begin{equation}
Y_{j}^{+}Y^{-}_{j} = (1+Y_{j-1})(1+Y_{j+1}),
\label{eqn:Y-system}
\end{equation}
representing a specialization of Eq.~\eqref{eqn:modified_Y} when $\lm_{j}\equiv 0$.
These relations have a purely group-theoretic origin and represent the kinematic input to the problem.
The physical input of the solution is encoded in analytic properties of $Y^{\rm D}_{j}$ in the complex plane.
In the present case consists of a double zero at the origin $\v=0$ and a pair of simple poles at 
$\pm (j+1)\tfrac{\ii}{2}$, while the large-$\v$ asymptotics matches that of the infinite-temperature Gibbs equilibrium, that is
$Y_{j}(\v)\sim j(j+2)$ as $|\v|\to \infty$.
The only physically relevant input lies inside the strip $\mathcal{P}$, where we find for every function $Y_{j}$
a double zeros at the origin, yielding the local TBA source terms
\begin{equation}
d_{j}(\v)=\log t(\v;0)^{2}.
\label{eqn:dimer_local_source}
\end{equation}
The TBA source terms obey the fusion property, $\square g_{j}=0$, where
\begin{equation}
g_{1}(\v) = \log\left(4\v^{2}(\v + \tfrac{\ii}{2})(\v - \tfrac{\ii}{2})\right).
\label{eqn:dimer_canonical_source}
\end{equation}

\paragraph{N\'eel state.}

The antiferromagnetic (N\'eel) state
\begin{equation}
\ket{\rm N} = \ket{\uparrow \downarrow}^{\otimes L/2},
\end{equation}
represents another particularly simple example which can be treated exactly (see e.g. \cite{WoutersGGE14,Brockmann14}).
The initial two $Y$-functions read explicitly~\cite{StringCharge}
\begin{eqnarray}
Y_{1}(\v) &=& \frac{4\v^{2}(12\v^{2}+19)}{(\v+\tfrac{\ii}{2})(\v-\tfrac{\ii}{2})(\v+\ii)(\v-\ii)},\\
Y_{2}(\v) &=& \frac{(\v+\tfrac{\ii}{2})(\v-\tfrac{\ii}{2})(2\v^{4}+7\v^{2}+2)}{\v^{2}(\v+\ii)(\v-\ii)(\v+\tfrac{3\ii}{2})(\v-\tfrac{3\ii}{2})},
\label{eqn:Neel_solution}
\end{eqnarray}
whereas the higher ones can be computed from the Y-system relations \eqref{eqn:Y-system}.
Importantly, the analytic data of $Y_{j}$ inside $\mathcal{P}$ are now given by double zeros [poles] for index $j$
being odd [even]. This implies
\begin{equation}
\g_{1}(\v) = \log\left(\frac{16\v^{2}}{(\v+\tfrac{\ii}{2})(\v-\tfrac{\ii}{2})}\right),
\label{eqn:Neel_canonical_source}
\end{equation}
while the higher source terms follow from the fusion condition,
\begin{equation}
\g_{j}=\sum_{k=1}^{j}\g^{[j+1-2k]}_{1}.
\label{eqn:fusion}
\end{equation}

It is worth stressing that the two examples given here in fact represent atypical initial conditions.
This is attributed to the fact that the TBA source have only components from $d^{\emptyset}_{j}$, implying $\lm_{j}\equiv 0$.
General initial states $\ket{\Psi}$ on the other hand involve non-vanishing $\lambda_{j}$ or, put differently,
are given by functions $\g_{j}$ which does violate the fusion condition \eqref{eqn:fusion}.
A practical drawback of this is that a full specification of a macrostate $\rho_{j}$ goes beyond the knowledge of
only $\g_{1}$.

\paragraph{Generic states.}

Here we analyse a non-trivial periodic state of $2$-spin ferromagnetic domain walls,
\begin{equation}
\ket{\Phi} = \ket{\uparrow \uparrow \downarrow \downarrow}^{\otimes L/4},
\label{eqn:domain_wall}
\end{equation}
representing one of the simplest generic initial states.
This state has already appeared in previous studies of quantum quench applications as the initial condition in 
refs.~\cite{FCEC14,PVC16}.
Expressions for the first few $Y$-functions can computed analytically, but are already rather formidable.
The initial $Y$-function for instance reads
\begin{equation}
Y_{1}(\v) = \frac{\left(64 \v^6+112 \v^4-20 \v^2+5\right) \left(192 \v^6+784 \v^4+1124 \v^2+491\right)}{\left(4 \v^2+1\right) \left(4 \v^2+5\right) \left(256 \v^8+1024 \v^6+1696 \v^4+1248
   \v^2+29\right)},
\end{equation}
with four zeros at $\pm(0.382\pm0.234\, \ii)$ and two poles at $\pm0.155\, \ii$, in $\mathcal{P}$.
It can furthermore be easily verified analytically that the universal form of the Y-system \eqref{eqn:Y-system} is now no longer 
satisfied (see also \cite{PVC16}). This specifically means that after subtracting the singular components $d^{\emptyset}_{j}$ from the 
TBA source terms $d_{j}$, we are left with non-vanishing components $s\star \lambda_{j}$ which in turn determine the node-dependent
terms $\lm_{j}$ in the modified Y-system relations \eqref{eqn:modified_Y}.

\section{Quench action}
The Quench Action approach~\cite{CE13} (see \cite{Caux16_review} for a review) has been introduced as a functional integral
approach towards time-evolution of local observables in thermodynamically large integrable many-body systems. The steady-state
limit of the evolution is entirely determined from the stationary part of the Quench Action functional
\begin{equation}
\mathcal{Z}_{\rm QA}=\int \mathcal{D}[\rho_{j}]
\exp{\Big(-L\sum_{j}\int_{\mathbb{R}}\dd \v(\g_{j}(\v)\rho_{j}(\v) + \mathfrak{s}_{j}(\v))\Big)},
\label{eqn:QA}
\end{equation}
where the state-dependent terms $\g_{j}$ represent the mode decomposition of the (density of the) logarithmic overlap
between $\ket{\Psi}$ and a general macrostate, namely
\begin{equation}
\sum_{j}\int_{\mathbb{R}}\dd \v\,\g_{j}(\v)\rho_{j}(\v) \equiv
\lim_{L\to \infty}L^{-1}\log |\braket{\Psi}{\{\v_{j}\}}|^{2}.
\label{eqn:logarithmic_overlap}
\end{equation}
Saddle-point evaluation of the functional $\mathcal{Z}_{\rm QA}$ yields a sought-for macrostate $\rho_{j}$.
As the choice of notation already suggests, $\g_{j}$ of Eq.~\eqref{eqn:logarithmic_overlap} are nothing but the
chemical potentials corresponding to individual particle modes.
This can be most easily be seen by comparing Eq.~\eqref{eqn:logarithmic_overlap} with the form of Eq.~\eqref{eqn:ensemble}.

\end{widetext}



\end{document}